\documentclass[prd,aps,showpacs,preprintnumbers,amssymb]{revtex4}
\usepackage{axodraw}
\usepackage{color}
\usepackage{epsf}

\def\e3p{$\eta \rightarrow 3 \pi$}

\begin{document}
\title{%
\hfill{\normalsize\vbox{%
\hbox{}
 }}\\
{Do we need Feynman diagrams for higher order perturbation theory?}}

\author{Renata Jora $^{\it \bf a}$~\footnote[1]{Email:
 rjora@theory.nipne.ro}}
 \affiliation{$^ {\bf \it a}$ National Institute of Physics and Nuclear Engineering, PO Box MG-6, Bucharest-Magurele, Romania.}
\date{\today}

\begin{abstract}
We compute the two loop and three loop corrections to the beta function for Yang Mills theories in the background gauge field method
and using the background gauge field as the only source. The calculations are based on the separation of the one loop effective potential into zero and positive modes
contributions and are entirely analytical. No two or three loop Feynman diagrams are considered in the process.
\end{abstract}
\maketitle
\section{Introduction}
The instanton approach for $SU(N)$ gauge theories with or without fermions has been initiated by 't Hooft \cite{Hooft} and further developed in \cite{Gross} and
\cite{Vainshtein}. In this method the separation of quantum degrees of freedom into zero modes (spin dependent) and positive modes (spin independent) is crucial. Moreover the so called zero modes have an " antiscreening " effect which is ultimately responsible for asymptotic freedom. The presence of fermions has an opposite effect. In \cite{Jora1} we suggest that in essence the magnetic properties of the QCD vacuum play a decisive role in the chiral symmetry breaking. Furthermore we show in \cite{Jora2} that in the process of gluino
decoupling from supersymmetric QCD separation into zero and positive modes is very important.

A very useful method for computing beta functions for the gauge coupling constant is the background gauge field method \cite{Abbott} which is based on the decomposition of the gauge field into a
background gauge field and a fluctuating field, the quantum gauge field. Even from the dawn of this method the background gauge field was regarded as an alternate source. However  the regular sources $J(x)$ and $\eta(x)$, $\eta'(x)$ (corresponding to the quantum gauge fields and ghost respectively) are introduced and one uses the conventional functional formalism to derive beta function or other loop corrections. The reason is simple; the background gauge field does not couple linearly to the other fields (as linear terms are canceled) and it is not obvious how one can compute simply Green functions with the background gauge field as a source.

In the present work we determine the two and three loop contributions to the beta function for Yang Mills theories using the background gauge field as the only source present in
the functional formalism. Of course the beta function is known up to the fourth order \cite{beta} in the MS scheme so our main interest lies in the method that we introduce and the possibility for that to be developed for higher orders. We rely on the well-known result of the one loop effective potential (derived either in the perturbative or in the instanton approach) and on the decomposition of the one loop operators into spin dependent and spin independent operators corresponding to each field. Our derivation is entirely based on an analytic functional approach  that does not involve the computation of any two or three loop Feynman diagrams.

\section{The one loop effective potential}

The Yang Mills Lagrangian in the background gauge field method (where the gauge field is separated into $B_{\mu}^a+ A_{\mu}^a$ and $B_{\mu}^a$ is the background gauge field) has the expression:
\begin{eqnarray}
{\cal L}&=&-\frac{1}{4g^2}[F^a_{\mu\nu}+D_{\mu}A^a_{\nu}-D_{\nu}A_{\mu}^a+f^{abc}A^b_{\mu}A^c_{\nu}]^2-
\nonumber\\
&-&\frac{1}{2g^2}(D^{\mu}A_{\mu}^a)^2+{\bar c}^a[(-D^2)^{ac}-D_{\mu}f^{abc}A_{\mu}^b]c^c
\label{mainl}
\end{eqnarray}

This lagrangian contains quantum gauge fields $A^a_{\mu}$ and ghosts $c^a$, ${\bar c}^a$ and can be separated into a quadratic contribution and a higher order one \cite{Peskin}.

The quadratic operator has the expression,
\begin{eqnarray}
{\cal L}_2=-\frac{1}{2g^2}[A^a_{\mu}(-(D^2)^{ac}g^{\mu\nu}-2f^{abc}F^{b \mu\nu})A^c_{\nu}]+{\bar c}^a(-(D^2)^{ac})c^c
\label{qadrmodes}
\end{eqnarray}
and can be decomposed as in:
\begin{eqnarray}
-\frac{1}{2g^2}[-(D^2)^{ac}-2f^{abc} F^{b\mu\nu}]
=-\frac{1}{2g^2}[-\partial^2+\Delta^{1}+\Delta^{2}+\Delta^J]=-\frac{1}{2g^2}\Delta,
\label{dec65748}
\end{eqnarray}

where $\Delta^1+\Delta^2$ is a spin independent operator and $\Delta^J$ is a spin dependent operator,
\begin{eqnarray}
\Delta^1&=&i[\partial^{\mu}B^b_{\mu}f^{abc}+B_{\mu}^bf^{abc}\partial^{\mu}]
\nonumber\\
\Delta^2&=&B^{a\mu}t^aB^b_{\mu}t^b
\nonumber\\
\Delta^j&=&-2f^{abc}F^{b\mu\nu}.
\label{res324}
\end{eqnarray}

No spin dependent operator acts on ghosts.

The one loop effective potential for a Yang Mills theory is obtained by computing,
\begin{eqnarray}
\exp[i\Gamma[B]]=\exp[i\int d^4 x [-\frac{1}{4g^2}(F^a_{\mu\nu})^2]][\det(\Delta_{G,1})]^{-1/2}\det(\Delta_{G,0}),
\label{det55}
\end{eqnarray}

where $\Delta_{G,1}$ refers to the gauge fields  and $\Delta_{G,0}$ to the ghost fields.
This leads to:
\begin{eqnarray}
\Gamma[B]=-\frac{1}{4}(\frac{1}{g^2}\int d^4 x (F^a_{\mu\nu})^2+\frac{1}{2}\ln[\det(\Delta_{G,1})]-\ln[\det(\Delta_{G,0})]).
\label{loop8989}
\end{eqnarray}

The logarithms are then expanded as in:
\begin{eqnarray}
&&\ln\det\Delta_{r,j}=\ln\det(-\partial^2)+ Tr[(-\partial^2)^{-1}(\Delta_1+\Delta_2+\Delta_J)-
\nonumber\\
&&-\frac{1}{2}(-\partial^2)^{-1}(\Delta_1+\Delta_2+\Delta_J)(-\partial^2)^{-1}(\Delta_1+\Delta_2+\Delta_J)+...].
\label{someexp657}
\end{eqnarray}



In Eq (\ref{someexp657}) the operator $(-\partial^2)^{-1}$ should be simply regarded as Feynman operators. For example,
\begin{eqnarray}
&&Tr [(-\partial^2)^{-1}(\Delta_1+\Delta_2+\Delta_J)(-\partial^2)^{-1}(\Delta_1+\Delta_2+\Delta_J)]\approx
\nonumber\\
&&\approx
Tr \int \int d^4x d^4y S(x-y)(\Delta_1+\Delta_2+\Delta_J)(x)S(y-x)(\Delta_1+\Delta_2+\Delta_J)(y),
\label{eq2131}
\end{eqnarray}

where $S(x-y)$ is the Feynman propagator.

 Each operator $\Delta$ has the decomposition from Eq (\ref{res324})with the following calculated contribution to the one loop effective potential:
\begin{eqnarray}
&&-\frac{1}{4g^2}\int d^4x (F^a_{\mu\nu})^2\longrightarrow -\frac{1}{4}[-4N\ln{\frac{M^2}{k^2}}]\int d^4x (F^a_{\mu\nu})^2\,\,\,{\rm zero\, modes\, contribution\,for\,quantum\,gauge\,fields}
\nonumber\\
&&-\frac{1}{4g^2}\int d^4x (F^a_{\mu\nu})^2\longrightarrow -\frac{1}{4}[\frac{2}{3}N\ln{\frac{M^2}{k^2}}]\int d^4x (F^a_{\mu\nu})^2\,\,\,{\rm positive\, modes\, contribution\,for\,quantum\,gauge\,fields}
\nonumber\\
&&-\frac{1}{4g^2}\int d^4x (F^a_{\mu\nu})^2\longrightarrow -\frac{1}{4}[-\frac{1}{3}N\ln{\frac{M^2}{k^2}}]\int d^4x (F^a_{\mu\nu})^2\,\,\,{\rm positive\, modes\, contribution\,for\,ghost\,fields}.
\label{contr657}
\end{eqnarray}

However it is more convenient for us to represent these results as,
\begin{eqnarray}
&&\ln[\det(\Delta^1+\Delta^2)_{G,1}]=\frac{1}{3}NX\int d^4x (F^a_{\mu\nu})^2
\nonumber\\
&&\ln[\det(\Delta^J)_{G,1}]=-2NX\int d^4x (F^a_{\mu\nu})^2
\nonumber\\
&&\ln[\det(\Delta^1+\Delta^2)_{G,0}]=\frac{1}{12}NX\int d^4x (F^a_{\mu\nu})^2.
\label{somelist767}
\end{eqnarray}
Here X is the regularized part of the one loop integral (see Eq (\ref{fin545}) in Section VI).

\section{The method}

In order to obtain  higher order corrections to the beta function we need to expand in the trilinear and quadrilinear terms in the Lagrangian. These are summarized below:

\begin{eqnarray}
{\cal L}_{3,4}&=&-\frac{1}{2g^2}(D_{\mu}A^a_{\nu}-D_{\nu}A^a_{\mu})f^{abc}A^{b\mu}A^{c\nu}-
\nonumber\\
&-&\frac{1}{4g^2}f^{abc}f^{ade}A_{\mu}^bA_{\nu}^cA^{d\mu}A^{e\nu}+{\bar c}^a(-D_{\mu}f^{abc}A_{\mu}^b)c^c
\label{therest54}
\end{eqnarray}

Then the effective action reduces to:
\begin{eqnarray}
e^{i \Gamma[B]}&=&\int {\cal D}A {\cal D}c\exp[i\int d^4x({\cal  L}+{\cal L}_{ct}]=
\nonumber\\
&=&\int {\cal D}A{\cal D}c\exp[i\int d^4 x[-\frac{1}{4g^2}(F^a_{\mu\nu})^2+{\cal L}_{ct}+{\cal L}_2+{\cal L}_{3,4}]]=
\nonumber\\
&=&\int {\cal D}A {\cal D}c \exp[i \int d^4x[-\frac{1}{4g^2}(F^a_{\mu\nu})^2+{\cal L}_2+{\cal L}_{ct}]]
\nonumber\\
&\times&[1+i\int d^4x {\cal L}_{3,4}-\frac{1}{2} \int \int d^4 x d^4 y{\cal L}_{3,4}(x){\cal L}_{3,4}(y)+....]
\label{exp6578}
\end{eqnarray}

where ${\cal L}_{ct}$ is the counterterm Lagrangian.

We plan to compute two or three loop contributions to the beta function using a  simple novel procedure which uses the background gauge field as the only source in the functional
approach.
For that we denote:
\begin{eqnarray}
&&{\cal L}_{4}=-\frac{1}{4g^2}f^{abc}f^{ade}A^b_{\mu}A^c_{\nu}A^{d\mu}A^{e\nu}
\nonumber\\
&&{\cal L}_{3a}=-\frac{1}{g^2}D_{\mu}A^a_{\nu}f^{abc}A^{b\mu}A^{c\nu}
\nonumber\\
&&{\cal L}_{3b}={\bar c}^a(-D_{\mu}f^{abc}A^b_{\mu})c^a.
\label{terms5454}
\end{eqnarray}

The two loop expansion contains the terms:
\begin{eqnarray}
&&i\int d^4x {\cal L}_4
\nonumber\\
&&-\frac{1}{2}\int \int d^4 x d^4y {\cal L}_{3a}(x){\cal L}_{3a}(y)
\nonumber\\
&&-\frac{1}{2} \int \int d^4 x d^4 y {\cal L}_{3b}(x){\cal L}_{3b}(y)
\nonumber\\
&&-\int \int d^4x d^4 y {\cal L}_{3a}(x){\cal L}_{3b}(y).
\label{twoloop97879}
\end{eqnarray}

The three loop corrections come from the terms:
\begin{eqnarray}
&&-\frac{1}{2}\int\int d^4x d^4 y {\cal L}_4(x){\cal L}_4(y)
\nonumber\\
&&-\frac{i}{2}\int\int \int d^4 x d^4 y d^4z {\cal L}_{3a}(x){\cal L}_{3a}(y){\cal L}_4(z)
\nonumber\\
&&-\frac{i}{2}\int\int \int d^4 x d^4 y d^4z {\cal L}_{3b}(x){\cal L}_{3b}(y){\cal L}_4(z)
\nonumber\\
&&-i\int\int \int d^4 x d^4 y d^4z {\cal L}_{3a}(x){\cal L}_{3b}(y){\cal L}_4(z)
\nonumber\\
&&+\frac{1}{24}\int\int\int\int d^4x d^4y d^4z d^4u {\cal L}_{3a}(x){\cal L}_{3a}(y){\cal L}_{3a}(z){\cal L}_{3a}(u)
\nonumber\\
&&+\frac{1}{24}\int\int\int\int d^4x d^4y d^4z d^4u {\cal L}_{3b}(x){\cal L}_{3b}(y){\cal L}_{3b}(z){\cal L}_{3b}(u)
\nonumber\\
&&+\frac{6}{24}\int\int\int\int d^4x d^4y d^4z d^4u {\cal L}_{3a}(x){\cal L}_{3a}(y){\cal L}_{3b}(z){\cal L}_{3b}(u)
\nonumber\\
&&+\frac{4}{24}\int\int\int\int d^4x d^4y d^4z d^4u {\cal L}_{3a}(x){\cal L}_{3a}(y){\cal L}_{3a}(z){\cal L}_{3b}(u)
\nonumber\\
&&+\frac{4}{24}\int\int\int\int d^4x d^4y d^4z d^4u {\cal L}_{3b}(x){\cal L}_{3b}(y){\cal L}_{3b}(z){\cal L}_{3a}(u).
\label{threeloop54645}
\end{eqnarray}

It turns out that in the expansion (\ref{twoloop97879}) and (\ref{threeloop54645}) of the lagrangian all combinations of terms that appear can be also decomposed into spin dependent and spin independent factors that can be obtained by suitable differentiation of the corresponding quadratic operators.

Before proceeding further we need to revise the rules of differentiation and integration in a general functional formalism and specifically for our case. Note that we are dealing only with gaussian integrals and their derivatives.  We start with the simple formula:
\begin{eqnarray}
\prod_k \int d\xi_k \exp[-\xi_i B_{ij}\xi_j]=
\prod_k\int dx_k \exp[-b_ix_i^2]=
\prod_i\sqrt{\frac{\pi}{b_i}}={\rm const}[\det{B}]^{-1/2}
\label{firstform}
\end{eqnarray}
In what follows we will drop the constant factors.

We extend this to a slightly more complicated case; assume the following:
\begin{eqnarray}
\prod_k \int   d \xi_k \exp[\xi_i B_{ij}\xi_j]\exp[\xi_i D_{ij} \xi_j]=\prod_i\sqrt{\frac{\pi}{\det(B+D)}}
\label{somfo7680}
\end{eqnarray}

Note that B and D correspond in our case to the spin dependent and spin independent operators respectively in the quadratic part of the Lagrangian.
Let us now differentiate one of the above factors with respect to a quantity $H_m$ where the index includes any type of subscript (In the end this $H_m$ will be the background gauge field tensor or any component of it).
\begin{eqnarray}
&&\prod_k \int d \xi_k \exp[-\xi_i B_{ij} \xi_j]\frac{\delta}{\delta H_m} \exp[-\xi_i D_{ij} \xi_j]=
\nonumber\\
&&=-\frac{1}{2}[\det(B+D)]^{-1/2} \sum_{i,j} \frac {\partial D_{ij}}{\partial H_m} (D+B)^{-1}_{ji}=-\frac{1}{2}[\det(B+D)]^{-1/2}\sum_i\frac{\delta d_i}{\delta H_m}\frac{1}{d_i+b_i}
\label{res333}
\end{eqnarray}

Here we assumed that the operators B and D are diagonalized by the same unitary operators in the functional formalism and that $b_i$ and $d_i$ are their eigenmodes.  This is true
provided that the one loop effective potential does not contain gauge invariants of order higher than two in the gauge tensor (see section IV for the proof).


We need to compute higher order derivatives of various types:
\begin{eqnarray}
&&\prod_k  \int d\xi_k \exp[-\xi_i B_{ij} \xi_j]\frac{\delta^2}{\delta H_m \delta H_n}\exp[-\xi_i D_{ij}\xi_j]=
\nonumber\\
&&[-\frac{1}{2}\frac{\delta^2 d_i}{\delta H_n \delta H_m}\frac{1}{b_i+d_i}+\frac{\delta d_i}{\delta H_m}\frac{\delta d_j}{\delta H_n}
(\frac{1}{4}\frac{1}{(b_i+d_i)(b_j+d_j)}+\frac{1}{2}\frac{1}{(b_i+d_i)^2}\delta_{ij})]
\nonumber\\
&&\times[\det(B+D)]^{-1/2}.
\label{rr4343}
\end{eqnarray}

Furthermore,

\begin{eqnarray}
&&\prod_k \int d \xi_k \frac{\delta}{\delta H_p}\exp[-\xi_i B_{ij} \xi_j]\frac{\delta^2}{\delta H_m \delta H_n} \exp[-\xi_i D_{ij} \xi_j]=
\nonumber\\
&&[\frac{\delta b_k}{\delta H_p}\frac{\delta^2 d_i}{\delta H_m \delta H_n}(\frac{1}{4}\frac{1}{(b_i+d_i)(b_j+d_j)}+\frac{1}{2}\frac{1}{(b_i+d_i)^2}\delta_{ik})+
\nonumber\\
&&\frac{\delta b_k}{\delta H_p}\frac{\delta d_i}{\delta H_m}\frac{\delta d_j}{\delta H_n}(-\frac{1}{8}\frac{1}{(b_i+d_i)(b_j+d_j)(b_k+d_k)}
-\frac{1}{4}\frac{1}{(b_i+d_i)^2(b_j+d_j)}\delta_{ik}
-\frac{1}{4}\frac{1}{(b_i+d_i)^2(b_j+d_j)}\delta_{jk}
\nonumber\\
&&-\frac{1}{4}\frac{1}{(b_i+d_i)^2(b_k+d_k)}\delta_{ij}+\frac{1}{2}\frac{1}{(b_i+d_i)^3}\delta_{ij}\delta_{ik})][\prod_l(b_l+d_l)]^{-1/2}.
\label{uyuy76}
\end{eqnarray}

We treat separately the ghost terms. Thus,
\begin{eqnarray}
\prod_k  \int d\theta_k d \theta_k^* \exp[\theta_i^* C_{ij} \theta_j]=
\prod_k \int d z_i dz_i^*  \exp[-c_i |z_i|^2]=\prod_i c_i=\det[C]
\label{gh456}
\end{eqnarray}

from which we can deduce,

\begin{eqnarray}
\prod_k \int d \theta_k d \theta_k^* \frac{\delta}{\delta H_m}\exp[\theta_i^* C_{ij} \theta_i]=
\prod_k \int dz_k dz^*_k \frac{\delta}{\delta H_m} \exp[-z_i^*c_i z_i]=
\frac{\delta c_i}{\delta H_m}\frac{1}{c_i}[\prod_k c_k]
\label{firste43526}
\end{eqnarray}

and further,

\begin{eqnarray}
&&\prod_k \int d \theta_k d \theta_k^* \frac{\delta^2}{\delta H_m \delta H_n}\exp[\theta_i^* C_{ij} \theta_i]=
\prod_k \int dz_k dz^*_k \frac{\delta^2}{\delta H_m \delta H_n} \exp[-z_i^*c_i z_i]=
\nonumber\\
&&\frac{\delta^2 c_i}{\delta H_m \delta H_n}\frac{1}{c_i} [\prod_k c_k]
+\frac{\delta c_i}{\delta H_m}\frac{\delta c_j}{\delta H_n}[-\frac{1}{c_i^2}\delta_{ij}+\frac{1}{c_ic_j}][\prod_k c_k].
\label{finform65789}
\end{eqnarray}

In general higher order derivatives will appear as one increases the order in perturbation theory. All of them can be easily computed along the same line.

We will assume for the moment that the one loop effective potential contains only terms proportional to the square of the gauge tensor. In order to better illustrate the method we compute
in detail the simplest contribution to the two loop beta function, that coming from the quadrilinear term in Eq(\ref{twoloop97879}):

\begin{eqnarray}
\int {\cal D} A \frac{-i}{4g^2}f^{abc}f^{ade}A_{\mu}^bA_{\nu}^cA^{d\mu}A^{e\nu}\exp[i\int d^4 x[-\frac{1}{4g^2}(F^a_{\mu\nu})^2+{\cal L}_2]]
\label{qop657}
\end{eqnarray}

The correct structure can be obtained from:
\begin{eqnarray}
&&\int {\cal D} A \frac{ig^2}{4}\int d^4 x d^4y \delta(x-y)\frac{\delta^2}{\delta F^b_{\mu\nu}(x) \delta F^{b\mu\nu}(y)}\exp[\int d^4 x \frac{i}{g^2}f^{abc}F^{b\mu\nu}A^a_{\mu}A^c_{\nu}]=
\nonumber\\
&&=\int {\cal D} A \frac{-i}{4g^2}f^{abc}f^{ade}A_{\mu}^bA_{\nu}^cA^{d\mu}A^{e\nu}\exp[\int d^4 x \frac{i}{g^2}f^{abc}F^{b\mu\nu}A^a_{\mu}A^c_{\nu}]
\label{res45367}
\end{eqnarray}

so it is clear that this operator comes only from the spin dependent part in the one loop effective potential.
We apply Eq (\ref{rr4343}). The contribution multiplying $\frac{\delta d_i}{\delta H_m}\frac{\delta d_j}{\delta H_n}$ is clearly coming only from
\begin{eqnarray}
\frac{\delta \exp[k_1\int d^4u (F^a_{\mu\nu})^2(u)]}{\delta F^b_{\rho\sigma}(x)}
\times\frac{\delta \exp[k_1\int d^4u (F^a_{\mu\nu})^2(u)]}{\delta F^{b \rho\sigma}(y)}=
4k_1^2 F^b_{\rho\sigma}F^{b\rho\sigma}
\label{res4353}
\end{eqnarray}

and depends only on the first term in the exponential.
The second type of terms are stemming from $\frac{\delta^2d_i}{\delta H_m \delta H_n}$ and they may involve besides the term similar to that in Eq(\ref{res4353}) another contribution:
\begin{eqnarray}
&&\int d^4 x d^4 y \delta(x-y)\frac{\delta^2}{\delta F^b_{\rho\sigma}(x)\delta F^{b \rho \sigma}(y)}[k_1\int d^4u (F^a_{\mu\nu})^2 (u)]
\times
\exp[k_1\int d^4u (F^a_{\mu\nu})^2(u)]
=
\nonumber\\
&&=\int d^4 x d^4 y \delta(x-y)[ 2k_1 \delta(x-y)]\times
\exp[k_1\int d^4u (F^a_{\mu\nu})^2(u)]
\label{res434232}
\end{eqnarray}

Let us show that terms of the type,
\begin{eqnarray}
\int d^4 x d^4 y \delta(x-y)[ 2k_1 \delta(x-y)]\times \exp[k_1\int d^4x F(x)^2]
\label{ex3232}
\end{eqnarray}

should be disregarded. For that we rewrite Eq(\ref{ex3232}) as,
\begin{eqnarray}
&&\int\int d^4x d^4y \delta(x-y)\frac{\partial^2 (Tr[\int d^4u d^4 v \Delta_J(u)S(u-v)\Delta_J(v)S(v-u)])}{\partial F^b_{\rho \sigma} \partial F^{b\rho\sigma}}\approx
\nonumber\\
&&\approx \int \int d^4x d^4 y \delta(x-y)S(x-y)S(y-x)=\int d^4 x S(x-x)S(x-x).
\label{u7686}
\end{eqnarray}
It is clear that this leads to bubble diagrams not connected to any external legs such that they do not contribute to the beta function. In what follows we will apply quite often this result especially for the three loop case where conveniently a number of terms will be dropped for this very reason. However we should note that there is no one to one correspondence between our approach and the standard functional formalism and one cannot just simply replace delta function by propagators in order to find the relation between the two of them.

Finally one finds the correct answer for the quadrilinear contribution.
\begin{eqnarray}
&&\int {\cal D} A \frac{-i}{4g^2}f^{abc}f^{ade}A_{\mu}^bA_{\nu}^cA^{d\mu}A^{e\nu}\exp[i\int d^4 x[-\frac{1}{4g^2}(F^a_{\mu\nu})^2+{\cal L}_2]]=
\nonumber\\
&&ig^2[-\frac{1}{8}\int d^4 x d^4 y \delta(x-y)\frac{\delta^2}{\delta F^b_{\mu\nu}(x)F^{b\mu\nu}(y)}\exp[-2NX\int d^4 z F^2(z)]\exp[\frac{5}{3}N X \int d^4 z F^2(z)]+
\nonumber\\
&+&\int d^4 x d^4 y \delta(x-y)\frac{3}{16}\frac{\delta}{\delta F^b_{\mu\nu}(x)}\exp[-2NX \int d^4 z F^2(z)]\times\exp[\frac{10}{3}N X\int d^4 z F^2(z)]\times
\nonumber\\
&&\times \frac{\delta}{\delta F^b_{\mu\nu}(x)}\exp[-2NX\int d^4 z F^2(z)]]\times
{\rm one\,loop\, contribution}
\nonumber\\
&=&ig^2 X^2N^2\int d^4 z F^2(z)\times{\rm one\,loop\, contribution}\times \exp[-\frac{1}{3}NX\int d^4 z F^2(z)][1+...].
\label{firsteerm55}
\end{eqnarray}

\section{Independence of operators}

In the previous section  we used heavily the fact that the operators B and D which correspond to the spin dependent and respectively spin independent quadratic operators in the lagrangian can be diagonalized by the same unitary operators. In what follows we will show that this assumption not only holds at one loop but it is also applicable in each order of perturbation theory.
At the one loop level this is evident from:
\begin{eqnarray}
&&Tr\ln[1+(-\partial^2)^{-1}(\Delta_1+\Delta_2+\Delta_J)]=
Tr\ln[1+(-\partial^2)^{-1}(\Delta_1+\Delta_2)]+Tr\ln[1+(-\partial^2)^{-1}(\Delta_J)]=
\nonumber\\
&&Tr\ln[1+(-\partial^2)^{-1}(\Delta_1+\Delta_2+\Delta_J)+(-\partial^2)^{-1}(\Delta_1+\Delta_2)(-\partial^2)^{-1}(\Delta_J)]
\label{eq2232}
\end{eqnarray}
where the contribution $(-\partial^2)^{-1}(\Delta_1+\Delta_2)(-\partial^2)^{-1}(\Delta_J)$ in the  expansion is zero if one consider only terms proportional to the square of the
gauge tensor.

It is quite safe to state that if the one loop effective potential contains only gauge invariants of order two (i.e. proportional to the square of the gauge tensor) the spin dependent and spin independent can be diagonalized by the same unitary matrix. Since the mixing of these operators can appear only from higher order gauge invariants  it is our task to show that higher order correction to the beta function cannot appear in our approach from these kind of terms (which in the Feynman diagram language correspond to one loop diagrams with more than two external legs).

In order to do that we will use as example Eq(\ref{uyuy76}) which, if we eliminate the assumption of simultaneous diagonalization, contains terms of the type:
\begin{eqnarray}
&&\frac{\partial B_{ij}}{\partial H_p}(-\partial^2 +B+D)^{-1}_{ji}\frac{\partial D_{kl}}{\partial H_m \partial H_n}(-\partial^2+B+D)_{lk}^{-1}
\nonumber\\
&&\frac{\partial B_{ij}}{\partial H_p}(-\partial^2 +B+D)^{-1}_{ji}\frac{\partial D_{kl}}{\partial H_m}(-\partial^2+B+D)_{lk}^{-1}\frac{\partial D_{rs}}{\partial H_n}(-\partial^2+B+D)_{sr}^{-1}
\label{pro978787}
\end{eqnarray}

Here $B=\Delta_1+\Delta_2$ and $D=\Delta_J$.   Then,
\begin{eqnarray}
&&(-\partial^2+\Delta_1+\Delta_2+\Delta_J)^{-1}=\frac{(-\partial^2)^{-1}}{1+(-\partial^2)^{-1}(\Delta_1+\Delta_2+\Delta_J)}=
\nonumber\\
&&=(-\partial^2)^{-1}[1-(-\partial^2)^{-1}(\Delta_1+\Delta_2+\Delta_J)+(-\partial^2)^{-1}(\Delta_1+\Delta_2+\Delta_J)(-\partial^2)^{-1}(\Delta_1+\Delta_2+\Delta_J)+...]
\label{res32345}
\end{eqnarray}

From Eq(\ref{uyuy76}) which must be of order $F^2$ in the gauge tensor one can deduce that the expansion (\ref{res32345}) should contains at most the second order term in $\Delta_1+\Delta_2+\Delta_J$. Let us assume that for example $(-\partial^2+B+D)^{-1}_{ji}$ contains the term $(\Delta_1+\Delta_2+\Delta_J)^2$. Then the other terms will be simply propagators and the second equation in (\ref{pro978787}) will become:
\begin{eqnarray}
Tr[\frac{\partial B}{\partial H_m}(-\partial^2)^{-1}((-\partial^2)^{-1}(\Delta_1+\Delta_2+\Delta_J))^2] \times
Tr[\frac{\partial D}{\partial H_n}](-\partial^2)^{-1}
 Tr[\frac{\partial D}{\partial H_p}](-\partial^2)^{-1}
\label{res32323}
\end{eqnarray}

But $Tr(D)=Tr(\Delta_1+\Delta_2)=Tr(\Delta_2)$  and in our derivation $H_m$ is just a component of the background gauge field. The corresponding term will thus not contribute since it is of an order higher than two in the gauge tensor. However in the first line of Eq (\ref{pro978787}) the factor $ Tr[\frac{\partial^2 D}{\partial H_n \partial H_p}]$ could contribute but it would lead to a delta function which in the full result gives a wrong space time structure (disconnected diagrams) (see Eq(\ref{u7686})).
 If on the other hand $(-\partial^2+B+D)^{-1}_{lk}$ contains the term $(\Delta_1+\Delta_2+\Delta_J)^2$ then we would obtain $Tr(B)=Tr(\Delta_J)=0$. Although illustrated for a particular case the result is quite general since these represent all types of terms that can appear.  This means that the inverse of the operator $(\partial^2+\Delta_1+\Delta_2+\Delta_J)^{-1}$ does not contain in its expansion  (\ref{res32345}) any term proportional to $B^2$ where B is the background gauge field. Then
 \begin{eqnarray}
 &&Q=1+(-\partial^2)^{-1}(\Delta_1+\Delta_2+\Delta_J)
 \nonumber\\
&&Q^{-1}=1-(\partial^2)^{-1}(\Delta_1+\Delta_J)
 \nonumber\\
 &&\det[Q]=\exp[Tr[(-\partial^2)^{-1}\Delta_2-\frac{1}{2}((\partial^2)^{-1}\Delta_1)^2-\frac{1}{2}((\partial^2)^{-1}\Delta_J)^2+...]]
 \nonumber\\
 &&\frac{d \det Q}{d B}=\det Q Tr[Q^{-1}\frac{ dQ}{dB}]=
 \nonumber\\
 && =\det Q Tr[(-\partial^2)^{-1}\frac{ d\Delta_2}{dB}-
 -(-\partial^2)^{-1}\Delta_1(-\partial^2)^{-1}\frac{ d\Delta_1}{d B}-(-\partial^2)^{-1}\Delta_J(-\partial^2)^{-1}\frac{ d\Delta_J}{d B}]
 \label{proof6575}
 \end{eqnarray}

Note that the differentiation of the operator in the third line leads to the result in the fifth line only and only if the one loop effective potential contains only terms
at most proportional to the square of the gauge tensor and not to higher order ones. But in this case the spin independent operator and the spin dependent one can be diagonalized simultaneously. This result is applicable in each order of perturbation theory and simplifies our calculations considerably.
From this point on we will not need anymore the analogy with the standard functional approach and we will proceed with our method to compute the two and three loop correction to the beta function.

\section{The trilinear pure gauge term}

This term corresponds to:

\begin{eqnarray}
&&-\frac{1}{2g^4}[\int d^4x (A^{a\nu}D_{\rho}A^c_{\nu}A^{m\rho}f^{acm})(x)\int d^4y (A^{d\mu}D_{\sigma}A^e_{\mu}A^{n\sigma}f^{den})(y)]\exp[i\int d^4 x[-\frac{1}{4g^2}(F^a_{\mu\nu})^2+{\cal L}_2]]
\label{optwo9786}
\end{eqnarray}

 It is simpler in this case to work with the gauge tensor $F^a_{\mu\nu}t^a$ where $t^a$ is the generator in the adjoint representation such that,
 \begin{eqnarray}
 Tr(F^a_{\mu\nu}t^a F^{a\mu\nu}t^a)=NF^a_{\mu\nu}F^{a\mu\nu}
 \label{s333}
 \end{eqnarray}

 and $B^{\mu}=b^{\mu c}t^c$.

 First we notice that the part of the term in Eq (\ref{optwo9786}) that contains covariant derivatives can be easily derived from the spin independent quadratic operator
 in accordance to:

 \begin{eqnarray}
 \frac{\delta[\int d^4 x \exp[-\frac{i}{2g^2}A^a_{\mu}(\Delta^1+\Delta^2)^{ac}A^c_{\nu}]]}{\delta (B^{\rho})_{ac}}=
 \frac{1}{g^2}A^a_{\mu}D_{\rho}A^c_{\nu}g^{\mu\nu}\exp[-\frac{i}{2g^2}\int d^4 x A^a_{\mu}(\Delta^1+\Delta^2)^{ac}A^c_{\nu}]
 \label{somef4567}
 \end{eqnarray}

 Then,
 \begin{eqnarray}
 &&\frac{\delta^2}{\delta B^{\rho}_{ac} \delta B^{\sigma}_{de}}\exp[-\frac{i}{2g^2}\int d^4 x A^a_{\mu}(\Delta^1+\Delta^2)^{ac}A^c_{\nu}]=
 \nonumber\\
 &&[\frac{-i}{g^2}A^a_{\nu}A^e_{\nu}\delta^{\rho\sigma}\delta_{cd}+
 \frac{1}{g^4}A^a_{\nu}D_{\rho}A^c_{\nu}A_{\mu}^dD_{\sigma}A^e_{\mu}]\times
 \nonumber\\
 &&\times\exp[-\frac{i}{2g^2}\int d^4 x A^a_{\mu}(\Delta^1+\Delta^2)^{ac}A^c_{\nu}].
 \label{double657}
 \end{eqnarray}

We need two more component gauge fields which can be simply obtained from the spin dependent operator.
 The desired result is finally obtained from:
 \begin{eqnarray}
&& \frac{\delta^2}{\delta B^{\rho}_{ac}(x)\delta B^{\sigma}_{de}(y)}\exp[-\frac{i}{2g^2}\int d^4 xA^a_{\mu}(\Delta^1+\Delta^2)A^c_{\nu}]
 \frac{\delta}{\delta F^{mn}_{\rho\sigma}(u)}\exp[\frac{1}{g^2}\int d^4 x A^a_{\mu}F^{\mu\nu}_{ac}A^c_{\nu}]=
 \nonumber\\
 &&=[-\frac{i}{g^4}A^{m\rho}A^{n\sigma}A^{a\nu}A^e_{\nu}\delta^{\rho\sigma}\delta^{\mu\nu}\delta_{cd}\delta(x-y)+
 \frac{1}{g^6}(A^{a\nu}D_{\rho}A^c_{\nu})(x)A^{m\rho}(u)(A^{d\mu}D_{\sigma}A^e_{\mu})(y)A^{n\sigma}(u)]\times
 \nonumber\\
 &&\times\exp[-\frac{i}{2g^2}\int d^4 xA^a_{\mu}(\Delta^1+\Delta^2)A^c_{\nu}+\frac{1}{g^2}\int d^4 x A^a_{\mu}F^{\mu\nu}_{ac}A^c_{\nu}].
 \label{almostf3245}
 \end{eqnarray}

 Eq(\ref{almostf3245}) has the correct structure except for the space time dependence. We will use a small artifice in order to correct that.
 First we use:
 \begin{eqnarray}
 &&\frac{\delta J(y)}{\delta J(x)}=\delta^4(x-y)
 \nonumber\\
 &&\frac{\delta J(y)}{\delta (\delta^4(x-y))}=J(x)
 \label{resa3333}
 \end{eqnarray}

 Then,

 \begin{eqnarray}
 \frac{\delta^2}{\delta(\delta(w_1-w_2))\delta F^{mn}_{\rho\sigma}(w_1)}\frac{1}{g^2}
 \int d^4 u d^4 v A^{m\rho}(u)F^{mn}_{\rho\sigma}(u)A^{n\sigma}(v)\delta(u-v)=\frac{1}{g^2}[A^{m\rho}(w_1)A^{n\sigma}(w_2)+A^{m\rho}(w_2)A^{n\sigma}(w_1)]
 \label{der454}
 \end{eqnarray}

 Since both terms in the last line of Eq(\ref{der454}) contribute equally to the result we add a factor of 1/2. This leads to:

 \begin{eqnarray}
 &&-\frac{1}{2g^4}[\int d^4x (A^{a\nu}D_{\rho}A^c_{\nu}A^{m\rho}f^{acm})(x)\int d^4y (A^{d\mu}D_{\sigma}A^e_{\mu}A^{n\sigma}f^{den})(y)]\exp[i\int d^4 x[-\frac{1}{4g^2}(F^a_{\mu\nu})^2+{\cal L}_2]]=
 \nonumber\\
 &&-\frac{1}{2}g^2\int d^4x d^4 y d^4u d^4v \delta(x-u)\delta(y-v)f^{acm}f^{den}\times[\frac{1}{2}\frac{\delta^2}{\delta (B^{\rho})_{ac}(x) \delta (B^{\sigma})_{de}(y)}\exp[-i\frac{1}{2g^2}\int d^4x A^a_{\nu}(\Delta^1+\Delta^2)A^c_{\nu}]\times
 \nonumber\\
 &&\times\frac{\delta^2}{\delta F^{mn}_{\rho\sigma}(u) \delta(\delta(u-v))}\exp[\frac{1}{g^2}\int d^4 x A^{m\rho}F^{mn}_{\rho\sigma}A^{n\sigma}]\times {\rm one\,loop\,ghost\, term}+
 \nonumber\\
 &&+i A^a_{\nu}A^e_{nu}A^{m\sigma}A^{n\sigma}\times {\rm one\,loop\,term}]
 \label{somres543}
 \end{eqnarray}

 Using the fact that the last term in Eq(\ref{somres543}) is proportional to the already computed quadrilinear term we get:

\begin{eqnarray}
&&-\frac{g^2}{4}\int d^4x d^4 y d^4u d^4v \delta(x-u)\delta(y-v)f^{acm}f^{den}\times
\nonumber\\
&&\frac{3}{4}\frac{\delta^2}{\delta (B^{\rho})_{ac}(x)\delta (B^{\sigma})_{de}(y)}\exp[\frac{N}{3}X\int d^4x F^2]\times
\nonumber\\
&&\times\frac{\delta^2}{\delta F^{mn}_{\rho\sigma}(u) \delta(\delta(u-v))}\times\exp[-2NX\int d^4x F^2]\times\exp[\frac{5}{3}NX\int d^4x F^2]\times{\rm ghost\,contribution}
\nonumber\\
&&=-2ig^2N^2X^2\int d^4x F^2(x)\times{\rm one \,loop\,contribution},
\label{fin6578}
\end{eqnarray}

and,

 \begin{eqnarray}
 &&-\frac{1}{2g^4}[\int d^4x (A^{a\nu}D_{\rho}A^c_{\nu}A^{m\rho}f^{acm})(x)\int d^4y (A^{d\mu}D_{\sigma}A^e_{\mu}A^{n\sigma}f^{den})(y)]\exp[i\int d^4 x[-\frac{1}{4g^2}(F^a_{\mu\nu})^2+{\cal L}_2]]=
 \nonumber\\
&& = -4ig^2N^2X^2\int d^4x F^2(x)\times{\rm one \,loop\,contribution}.
\label{som77868}
\end{eqnarray}

\section{Terms that include ghosts}

 There is one quadratic term which contains ghosts and two higher order contributions.
  We will need to determine two terms, respectively:
  \begin{eqnarray}
&&-\frac{1}{2}( D_{\mu}{\bar c}^a f^{abc} A^b_{\mu}c^c)^2\exp[i\int d^4 x[-\frac{1}{4g^2}(F^a_{\mu\nu})^2+{\cal L}_2]]
\nonumber\\
 &&-\frac{1}{g^2}D_{\mu}{\bar c}^a f^{abc} A^b_{\mu}c^c(D_{\rho}A_{\sigma}^df^{def}A^{e\rho}A^{f\sigma})\exp[i\int d^4 x[-\frac{1}{4g^2}(F^a_{\mu\nu})^2+{\cal L}_2]].
 \label{ter4356}
 \end{eqnarray}

 We start by analyzing the first term in Eq(\ref{ter4356}).

Both these expressions contain the ghost fields mixed with quantum gauge fields. For the sake of simplicity we write:
 \begin{eqnarray}
 {\bar c}^a(-D^{\mu}f^{abc}A^b_{\mu})c^c \equiv
 D_{\mu}{\bar c}^a f^{abc} A^b_{\mu}c^c
 \label{gh5467}
 \end{eqnarray}
which is true up to a total derivative.
Moreover the quadratic term must also be written in a similar manner as:
 \begin{eqnarray}
 {\bar c}^a(-D^2)_{ac}c^c \equiv
 (-D^2{\bar c}^ac^c)
 \label{e32456}
 \end{eqnarray}

We can switch in all these terms the order of the ghost field without problem since we are dealing with the square of the trilinear operator. Then the analogy with the previous case is obvious and with exactly the same derivation we obtain:

\begin{eqnarray}
 &&-\frac{1}{2g^4}[\int d^4x (D_{\rho}{\bar c}^cc^aA^{m\rho}f^{acm})(x)\int d^4y(D_{\sigma}{\bar c}^ec^dA^{n\sigma}f^{den})(y)]\exp[i\int d^4 x[-\frac{1}{4g^2}(F^a_{\mu\nu})^2+{\cal L}_2]]=
 \nonumber\\
&&=-\frac{g^2}{2}\int d^4x d^4 y d^4u d^4v \delta(x-u)\delta(y-v)f^{acm}f^{den}\times
\nonumber\\
&&(-\frac{1}{8})\frac{\delta^2}{\delta (B^{\rho})_{ac}(x)\delta (B^{\sigma})_{de}(y)}\exp[\frac{1}{12}N X\int d^4x F^2]\times
\nonumber\\
&&\times\frac{\delta^2}{\delta F^{mn}_{\rho\sigma}(u)}\exp[-2NX\int d^4x F^2]\times
\nonumber\\
&&\times \exp[(\frac{5}{3}N-\frac{1}{12}N)X\int d^4F^2]\times{\rm one\,loop\,spin\,independent\,gauge\,contribution}
\nonumber\\
&&=ig^2\frac{1}{6}N^2X^2\int d^4x F^2(x)\times\exp[-\frac{1}{3}NX\int d^4F^2]\times{\rm one \,loop\,contribution}.
\label{fin657888}
\end{eqnarray}

In this approach the second term in Eq (\ref{ter4356}) will give no contribution since it will appear as a product of three functional derivatives corresponding to the spin dependent, spin independent and  ghost terms in the one loop potential and this would lead to a result proportional to a gauge invariant (in the background gauge field) of an order higher than two.

\section{Two loop beta function}

We add the results from Eq(\ref{firsteerm55}), Eq (\ref{fin6578}) and Eq(\ref{fin657888}) to obtain for the second order correction:
\begin{eqnarray}
-ig^2\frac{17}{6}N^2 X^2\int d^4 x(F^a_{\mu\nu})^2
\label{rez3232}
\end{eqnarray}

 Here X is just the result of the regularization at one loop. After taking into account all gauge and internal indices X amounts to a one loop scalar integral so one can write schematically for
the proper loop result:
\begin{eqnarray}
\approx \int d^4 x d^4 y F_{\mu\nu}^a U(x-y)F^{a\mu \nu} (y)
\label{reg554}
\end{eqnarray}

The two loop expression then corresponds to:
\begin{eqnarray}
\approx \int d^4 x d^4 y F_{\mu\nu}^a (U U)(x-y)F^{a\mu \nu} (y)
\label{res4343}
\end{eqnarray}

where $(UU)(x-y)$ is the result of the scalar two loop diagram with two bubbles and two external legs. But this regularized is just the square of U regularized at one loop so practically we do not need it. So finally we will take for X the expression:
\begin{eqnarray}
X=i\frac{1}{(4\pi)^2}\int_0^1\ln(\frac{x\Lambda^2}{-x(1-x)k^2})=i\frac{1}{(4\pi)^2}(1+\ln{\Lambda^2/k^2})
\label{fin545}
\end{eqnarray}
We multiply by a loop factor $\frac{1}{2}$ to obtain the second order contribution to the coupling constant:
\begin{eqnarray}
i\frac{g^2(k)}{4}\int
d^4 x (F_{\mu\nu}^a)^2 =i\frac{g^2}{4}[1-\frac{11N}{3}\frac{1}{(4\pi)^2}\ln{M^2/k^2}-\frac{34}{3}N^2\frac{g^2}{(4\pi)^4}\ln{M^2/k^2}+...]\int d^4 x (F_{\mu\nu}^a)^2
\label{y677}
\end{eqnarray}

From that the known result for the two loop beta function is obtained:
\begin{eqnarray}
\beta(g^2)=\frac{g^4}{(4\pi)^2}[-\frac{11}{3}N-\frac{34}{3}N^2\frac{g^2}{16\pi^2}].
\label{res43332}
\end{eqnarray}
Here we defined $\beta(g)=\frac{d g^2}{d\ln(\mu^2)}$.

\section{Three loop beta function}

We need to evaluate each term in the list of Eq (\ref{threeloop54645}).  We illustrate our work in some detail for the second one,
\begin{eqnarray}
&&-\frac{i}{2}\int \int \int d^4 xd^4 y d^4 z {\cal L}_{3a}(x) {\cal L}_{3a}(y) {\cal L}_4(z)=
\nonumber\\
&&=-\frac{i}{8}\int\int\int d^4x d^4 y d^4 z (A^{a\nu}D_{\rho}A^c_{\nu}A^{m\rho})(x)(A^{d\mu}D_{\sigma}A^e_{\mu}A^{n\sigma})(y)
f^{acm}f^{den}(-f^{prs}f^{pqt}A^r_{\alpha}A^s_{\beta}A^{q\alpha}A^{t\beta}).
\label{res4343}
\end{eqnarray}

and only list the results for the others.
The right structure in Eq(\ref{res4343}) can be obtained from:
\begin{eqnarray}
&&\int \int \int \int \int d^4x d^4 y d^4 zd^4 u d^4 v \delta(x-u) \delta(y-v)\frac{1}{2}f^{acm}f^{den}\frac{\delta^3}{\delta (B^{\rho})_{ac}(x)\delta(B^{\sigma})_{de}(y)}\exp[\frac{-i}{2g^2} \int d^4 x A^a_{\mu}(\Delta_1+\Delta_2)_{ac}A^{c\nu}]\times
\nonumber\\
&&\times\frac{\delta^4}{\delta (F^{\rho\sigma})_{mn}(v)\delta(\delta(u-v))\delta F^p_{\alpha\beta}(z)\delta F^{p\alpha\beta}(z)}\exp[\frac{i}{g^2}f^{abc}F^{b\mu\nu}A^a_{\mu}A^c_{\nu}]=
\nonumber\\
&&=\int \int \int \int \int d^4x d^4 y d^4 zd^4 u d^4 v \delta(x-u) \delta(y-v)f^{acm}f^{den}(A^{a\nu}D_{\rho}A^c_{\nu})(x)A^{m\rho}(u)(A^{d\mu}D_{\sigma}A^e_{\mu})(y)A^{n\sigma}(v)\times
\nonumber\\
&&\times
(-f^{prs}f^{pqt}A^r_{\alpha}A^s_{\beta}A^{q\alpha}A^{t\beta})(z)\times{\rm one\,loop\, gauge\, contribution}+
\nonumber\\
&&+\int \int \int \int \int d^4x d^4 y d^4 zd^4 u d^4 v \delta(x-u) \delta(y-v)(-i) f^{acm}f^{den}\delta_{cd}A^{a\nu}(x)A^e_{\nu}(x)\delta(x-y)A^{m\rho}(u)A^{n\sigma}(y)(v)\times
\nonumber\\
&&\times(-f^{prs}f^{pqt}A^r_{\alpha}A^s_{\beta}A^{q\alpha}A^{t\beta})(z)\times{\rm one\,loop\, gauge\, contribution}.
\label{contr65757}
\end{eqnarray}

The last term in Eq. (\ref{contr65757}) is proportional to $-\frac{1}{2}\int\int d^4 xd^4 y {\cal L}_4{\cal L}_4$ and needs to be subtracted from the result
whereas the first one is exactly what appears in Eq (\ref{res4343}).
Then a simple computation yields:
\begin{eqnarray}
-\frac{i}{2}\int \int\int d^4x d^4 y d^4 z {\cal L}_{3a}{\cal L}_{3a}{\cal L}_4=
12N^3X^3g^4+4N^3X^3g^4
\label{firs565757}
\end{eqnarray}

where the first quantity in the last line corresponds to the second term in Eq(\ref{contr65757}) and the second one is the subtracted contribution.

For completitude all contributions  are listed below:
\begin{eqnarray}
&&-\frac{1}{2}\int\int d^4x d^4 y {\cal L}_4(x){\cal L}_4(y)=-N^3X^3g^4
\nonumber\\
&&-\frac{i}{2}\int\int \int d^4 x d^4 y d^4z {\cal L}_{3a}(x){\cal L}_{3a}(y){\cal L}_4(z)=12N^3X^3g^4+4N^3X^3g^4
\nonumber\\
&&-\frac{i}{2}\int\int \int d^4 x d^4 y d^4z {\cal L}_{3b}(x){\cal L}_{3b}(y){\cal L}_4(z)=-\frac{1}{3}N^3X^3g^4
\nonumber\\
&&-i\int\int \int d^4 x d^4 y d^4z {\cal L}_{3a}(x){\cal L}_{3b}(y){\cal L}_4(z)=0
\nonumber\\
&&+\frac{1}{24}\int\int\int\int d^4x d^4y d^4z d^4u {\cal L}_{3a}(x){\cal L}_{3a}(y){\cal L}_{3a}(z){\cal L}_{3a}(u)=-(24+\frac{8}{9})N^3X^3g^4
\nonumber\\
&&+\frac{1}{24}\int\int\int\int d^4x d^4y d^4z d^4u {\cal L}_{3b}(x){\cal L}_{3b}(y){\cal L}_{3b}(z){\cal L}_{3b}(u)=(-\frac{1}{432}+\frac{1}{18})N^3X^3g^4
\nonumber\\
&&+\frac{6}{24}\int\int\int\int d^4x d^4y d^4z d^4u {\cal L}_{3a}(x){\cal L}_{3a}(y){\cal L}_{3b}(z){\cal L}_{3b}(u)=(\frac{1}{6}+\frac{2}{3})N^3X^3g^4
\nonumber\\
&&+\frac{4}{24}\int\int\int\int d^4x d^4y d^4z d^4u {\cal L}_{3a}(x){\cal L}_{3a}(y){\cal L}_{3a}(z){\cal L}_{3b}(u)=0
\nonumber\\
&&+\frac{4}{24}\int\int\int\int d^4x d^4y d^4z d^4u {\cal L}_{3b}(x){\cal L}_{3b}(y){\cal L}_{3b}(z){\cal L}_{3a}(u)=0.
\label{finalresultsthreeloops435}
\end{eqnarray}

We add the results in Eq(\ref{finalresultsthreeloops435}), introduce them in the effective potential, substitute for the value of X (see Eq(\ref{fin545})), divide by a factor of three (as corresponding to a second order differentiation of a product of three factors). This yields the following three loop beta function:
\begin{eqnarray}
\beta(g^2)=\frac{g^4}{(4\pi)^2}[-\frac{11}{3}N-\frac{34}{3}N^2\frac{g^2}{16\pi^2}-\frac{4033}{108}N^3\frac{g^4}{256 \pi^4}]
\label{res43332}
\end{eqnarray}

It is well known that the first two orders of the beta function are renormalization scheme independent whereas the higher orders can take a large number of values depending on the renormalization scheme.
  Our result  of $\frac{4033}{108}=37.3$ does not coincide, as expected,  with the standard MS result $\frac{2857}{54}=54.7$. There are no other three loop estimates in the literature except for the all orders beta function proposed by Pica and Sannino \cite{Sannino}:

\begin{eqnarray}
\beta (g^2)=-\frac{11}{3}\frac{g^4}{16\pi^2}\frac{N}{1-\frac{g^2}{8\pi^2}\frac{17}{11}N}.
\label{pica4535}
\end{eqnarray}

The corresponding three loop coefficient can be deduced to be $\frac{1156}{33}=35$ which is very close to our value.

\section{Discussion}

It is important to know the beta function for non-abelian gauge theories for several reasons. First the one loop coefficient of beta function was the main clue that these theories are endowed with asymptotic freedom. Second higher order coefficients can reveal information about the phase structure of these type of models. And it is always useful to learn more about the mathematical structures that lie at the basis of contemporary particle physics.

In the present work we introduced a new method for computing beta functions for Yang Mills theories and applied it for determining two and three loop corrections. Our approach relies entirely on the functional formalism and on the separation of the one loop operators onto spin dependent and spin independent ones. It turns out that each term in the expansion of the action can be decomposed in factors derived from one or another of these operators. The calculations are based almost entirely on functional differentiation and do not involve any two or three loops Feynman diagrams.

\section*{Acknowledgments} \vskip -.5cm
I am happy to thank J. Schechter for support and encouragement and for useful comments on the manuscript.
This work has been supported by PN 09370102/2009.

\appendix

\section{}

It is useful to give in what follows some results regarding the functional derivatives of the square of the gauge tensor (Here $F^a_{m\nu}t^a=F_{\mu\nu}$, where $t^a$ is the generator
in the adjoint representation).
\begin{eqnarray}
&&\frac{\delta^2}{\delta B^{\rho}_{ac}(x) \delta B^{\sigma}_{de}(y)}[\int d^4 z Tr(F_{\mu\nu})^2(z)]=
\nonumber\\
&&\int d^4 z 2[(F_{\mu\nu})_{gh}(z)\frac{\delta^2 (F^{\mu\nu})_{hg}}{\delta B^{\rho}_{ac}(x) \delta B^{\sigma}_{de}(y)}+
\frac{\delta (F_{\mu\nu})_{gh}(z)}{\delta B^{\rho}_{ac}(x)}\frac{\delta (F^{\mu\nu})_{hg}(z)}{\delta B^{\sigma}_{de}(y)}]=
\nonumber\\
&&8i[(F_{\rho\sigma})_{ae}(x)\delta_{cd}\delta(x-y)-(F_{\rho\sigma})_{dc}\delta_{ae}\delta(x-y)].
\label{someusefrs56}
\end{eqnarray}

Furthermore from this one can deduce:
\begin{eqnarray}
&&\int d^4x d^4 y d^4u d^4v \delta(x-u)\delta(y-v)f^{acm}f^{den}\times
\nonumber\\
&&\frac{\delta^2}{\delta (B^{\rho})_{ac}(x)\delta (B^{\sigma})_{de}(y)}\exp[k_1\int d^4z (F^a_{\mu\nu}F^{a\mu\nu})(z)]\times
\nonumber\\
&&\times\frac{\delta^2}{\delta F^{mn}_{\rho\sigma}(u) \delta(\delta(u-v))}\exp[k_2\int d^4z (F^a_{\mu\nu}F^{a\mu\nu})(z)]=
\nonumber\\
&&=-16ik_1k_2\int d^4 x (F_{\mu\nu}^a)^2\times\exp[(k_1+k_2)\int d^4z (F^a_{\mu\nu}F^{a\mu\nu})(z)]
\label{fin657899}
\end{eqnarray}

We also used the relations,
\begin{eqnarray}
&&Tr [t^at^b]=N\delta_{ab}
\nonumber\\
&& Tr[t^ct^at^ct^d]=\frac{N^2}{2}\delta_{ad}
\nonumber\\
&& -Tr[t^rt^ct^at^st^at^c]=\frac{N^3}{4}\delta_{rs}.
\label{usef3435}
\end{eqnarray}

Here all generators are in the adjoint representation and summation over repeated indices is understood.

\end{document}